# Negative Capacitance Tunnel Field Effect Transistor: A Novel Device with Low Subthreshold Swing and High ON Current

Nadim Chowdhury, S. M. Farhaduzzaman Azad and Quazi D.M. Khosru

Department of Electrical and Electronic Engineering

Bangladesh University of Engineering and Technology, Dhaka-1000, Bangladesh

In this paper we propose a modified structure of TFET incorporating ferroelectric oxide as the complementary gate dielectric operating in negative capacitance zone, called the Negative Capacitance Tunnel FET (NCTFET). The proposed device effectively combines two different mechanisms of lowering the sub threshold swing (SS) for a transistor garnering a further lowered one compared to conventional TFET. A simple yet accurate analytical tunnel current model for the proposed device is also presented here. The developed analytical model demonstrates high ON current at low $V_{GS}$ and exhibits lower SS.

## Introduction

If the empirical Moore's law holds forever, then in near future the dissipative power density in a chip will surpass the maximum heat removal limit, 1000 W/cm$^2$ (1) from an IC by using conventional technology. To reduce the power hunger of IC the transistors should be designed such that it can turn on at a very low voltage having significantly low sub-threshold slope with high $I_{ON}/I_{OFF}$ ratio. To maintain a low power density, the power supply voltage $V_{DD}$ has to go down with the scaling of device dimensions. But the mere reduction of the $V_{DD}$ reduces the ON current $I_{ON}$, so to meet up the $I_{ON}$ requirement the threshold voltage, $V_{TH}$ should be scaled with the scaling of the supply voltage. However, $I_{OFF}$ exponentially increases with the threshold voltage reduction, since $I_{OFF} \propto 10^{-V_{TH}/SS}$. So, to limit $I_{OFF}$ while maintaining a satisfactory $I_{ON}$ the sub-threshold slope (SS) has to go down. However, for the conventional MOSFET structure the SS cannot be reduced below 60mV/decade at room temperature. In order to circumvent this limit a number of novel devices has recently been reported to the literature such as Impact Ionization-MOSFET (2), NanoelectromechanicalFET (3), Suspended Gate MOSFET (4), Tunneling FET (5), Fe-FET (6). TFET can supersede the fundamental SS limitation of conventional MOSFET by the exploitation of band to band tunneling mechanism. Salahuddin and Datta (6) theoretically demonstrated that a ferroelectric (FE) insulator operating in the negative capacitance region could act as a step-up transformer of the surface potential in Metal Oxide Semiconductor structure, opening a new way for the realization of transistors with steeper subthreshold characteristics (SS < 60 mV/decade) without changing the basic physics of the FET. In this work we propose a novel device and present a theoretical framework to characterize the proposed device. Our device integrates the aforementioned two physical phenomena to obtain two fold reduction in the subthreshold slope.

## Device Structure

Verhulst et. al. (7-8) have shown that for TFET to work properly gate does not need to extend over the entire intrinsic region. Ref. (8) proposed a new type of TFET structure where the gate is aligned fully on top of the source region as shown in Fig.1 (a). In this device band to band tunneling occurs in the direction orthogonal to the gate. The structure of the proposed device is shown in Fig. 1(b). Traditional double gate Line TFET with the high-κ dielectric is the basis of this proposition. FE dielectric is assumed to be deposited on a commensurate metallic layer grown on high-κ oxide. This back to back oxide structure is chosen following the experimental demonstration (10). Fig.1 (b) exhibits the ntype-NCTFET where p+ region, intrinsic region and n+ region performs the role of source, channel and drain respectively. When a gate voltage $V_{GS}$ is applied it gets amplified by the negative capacitance action of the ferroelectric gate oxide and a higher surface potential appears in the highly doped source region as a result the semiconductor region underneath the gate becomes depleted until the appearance of inversion layer. At sufficiently high gate bias tunneling current emerges and continues to increase with the enhancement of gate voltage.

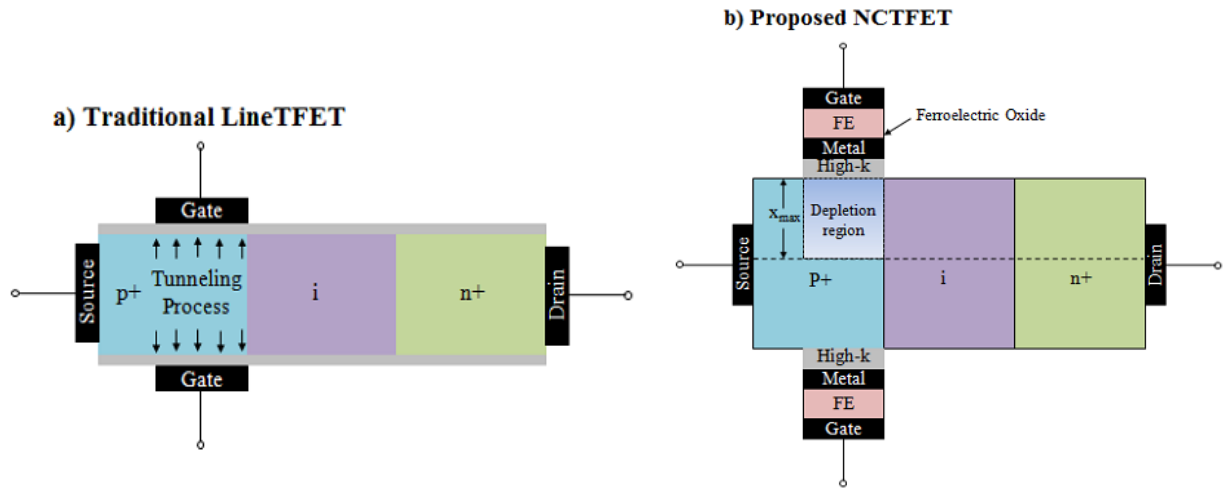

Figure 1. (a) The TFET structure that was modeled in Ref. (9), the arrows in the highly doped source region shows the direction of band to band tunneling. (b) Simplified structure of proposed new device Negative Capacitance Tunnel FET (NCTFET).

## Model Development

Our model is based on the 1-D electrostatic potential variation due to the application of gate voltage. Fig. 2 shows the 1-D capacitance model from the top gate to the semiconductor node in the source region along with the voltage drop across different capacitances. First, a tunnel drain current model is developed using the conventional TFET formalism which gives $I_{DS}-V_{GI}$ characteristics (11). Then, the voltage drop across FE material $V_{ins}$ is modeled to find $V_{GS}$, thus completing the current ($I_{DS}$–$V_{GS}$) model for the proposed NCTFET structure.

Modeling $I_{DS}$ – $V_{GI}$ characteristics

Careful investigation of NCTFET structure reveals that below FE material the structure is no different than conventional TFET. So, $I_{DS}$ –$V_{GI}$ characteristics can be modeled following the methodology presented in the literature. In the absence of BTBT a small current exists in the reverse biased TFET structure, which is called OFF current. But when the gate voltage is applied BTBT starts dominating over the OFF current. Carrier generation rate by BTBT process is described by Kanes Model (11) which provides the number of carriers tunneling from the valance band to the conduction per unit volume.

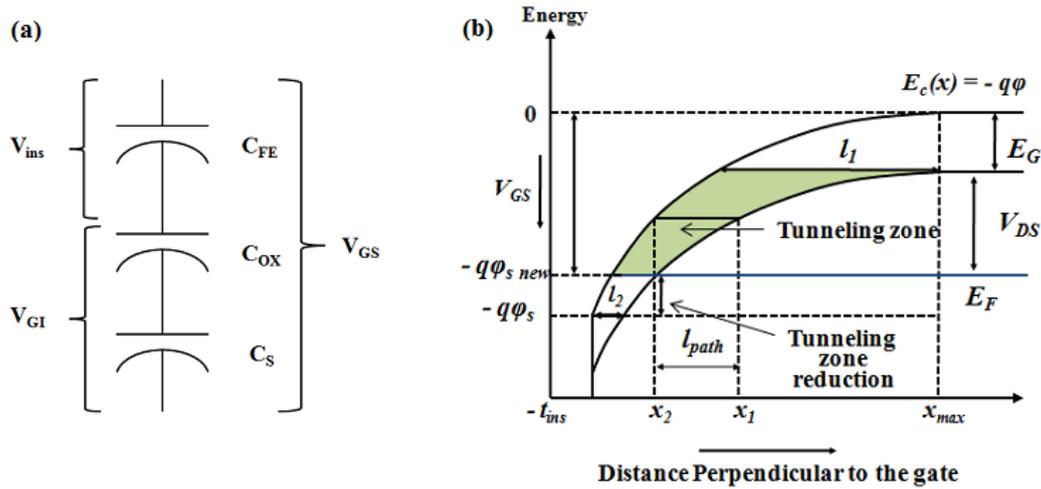

Figure 2. (a) 1-D capacitance model of NCTFET demonstrating the voltage drop across them. (b) Energy band diagram perpendicular to the gate illustrating the BTBT zone reduction at low $V_{DS}$

$$G = A \frac{E^D}{\sqrt{E_g}} \exp(-B \frac{E_g^{3/2}}{E}) \qquad [1]$$

Where E is the local electric field, Eg band gap of the material, A and B are material dependant parameter and D is the parameter, distinguishing the direct (D = 2) from the indirect (D = 2.5) tunneling process. So the source drain current of a convention TFET is given by the following equation.

$$I_{DS} = q \int G dv \qquad [2]$$

Where q is the unit charge and $v$ is the three dimensional volume inside the device where the tunneling phenomena take place. In our case, due to the one dimensional variation of electric field the generated Drain source can be written as

$$I_{DS} = qWL \int G\,dx \qquad [3]$$

This paper follows the formalism proposed by Vandenberghe et. al. (9) which utilizes depletion layer approximation neglecting the effect of inversion region and instead of inserting the local electric field, this paper utilizes the average electric field over the tunnel path into the generation rate. According to Ref (8) drain tunnel current $I_{DS}$ without including the impact of drain voltage $V_{DS}$ with respect to internal gate voltage $V_{GI}$ is given by the following equation.

$$I_{DS} = -\frac{qAL_{gs}WE_g^{D-1}}{2q^{D-1}} \int_{l_1}^{l_2} C_3 \frac{1}{l_{path}^{D-1}}\left(1 - \frac{C_1}{l_{path}^2}\right)\exp(-C_2 l_{path})\,dl_{path} \qquad [4]$$

$$\text{Where, } C_1 = \frac{2E_g \varepsilon_s}{q^2 N_a} \qquad [5]$$

$$C_2 = qB\sqrt{E_g} \qquad [6]$$

$$C_3 = \frac{qN_a}{\varepsilon_s l_{path}}\left(\frac{l_{path}^2}{2} + \frac{E_g \varepsilon_s}{q^2 N_a}\right) \qquad [7]$$

$$l_1 = \sqrt{C_1} \qquad [8]$$

$$l_2 = \sqrt{\frac{2\varepsilon_s}{qN_a}}\left(\sqrt{\varphi_s} + \sqrt{\varphi_s - \frac{E_g}{q}}\right) \qquad [9]$$

Where $L_{gs}$ is the gate source overlap length, W is the width of the device and $\varphi_s$ is the total band bending. Now if we assume that exponent part is varying is more rapidly than the polynomial factor upon the variation of path then we can find the current equation in a compact form.

$$I_{DS} = -\frac{AL_{gs}WE_g^{D-\frac{3}{2}}}{2q^{D-1}B}(H(l_2) - H(l_1)) \qquad [10]$$

$$\text{With } H(l) = C_3 \frac{1}{l^{D-1}}\left(1 - \frac{C_1}{l^2}\right)\exp(-C_2 l) \qquad [11]$$

According to Fig.3 $l_1$ is the longest tunnel path and $l_2$ is the shortest one, so BTBT probability increases from $l_1$ to $l_2$. That is why to find the total current with infinite $V_{DS}$ the integrand of eqn. (4) is integrated over the limit from $l_2$ to $l_1$. Now, to complete the on current model of the proposed device, we need to find the relation between the total band bending $\varphi_s$ and internal gate voltage $V_{GI}$. From the basic MOS physics we can write the following equation for $\varphi_s$.

$$\varphi_s = \left(\sqrt{t_{eff,ox}^2 \frac{qN_a}{2\varepsilon_s} + (V_{GI} - V_{FB})} - \sqrt{t_{eff,ox}^2 \frac{qN_a}{2\varepsilon_s}}\right)^2 \qquad [12]$$

With, $t_{eff,ox} = t_{ox}(\varepsilon_s)/(\varepsilon_{ox})$ where $\varepsilon_{ox}$ is the dielectric constant of the gate oxide and $\varepsilon_s$ is the dielectric constant of the silicon.

## Modeling $V_{ins}$

The top and the bottom metal plate across the FE material forms a MFEM capacitor ($C_{FE}$). The electrostatic dynamics of this negative capacitance $C_{FE}$ can be well described by LK (Landau-Khalatnikov) equation (12-13).

$$\rho \frac{d\bar{P}}{dt} + \bar{\nabla}_p U = 0 \qquad [13]$$

Where, U is the Gibbs free energy. We can express it in terms of anisotropy energy.

$$U = \alpha P^2 + \beta P^4 + \gamma P^6 - \bar{E}_{ext} \cdot \bar{P} \qquad [14]$$

Here, P is the polarization charge per unit area. At steady state operation dP/dt≈ 0 and for FE material P ≈ Q. In this case, the external electric field is given by $E_{ext} = V_{ins}/t_{ins}$. Taking all these into consideration and mingling (13) and (14) we get

$$V_{ins} = 2\alpha t_{ins} Q + 4\beta t_{ins} Q^3 + 6\gamma t_{ins} Q^5 \qquad [15]$$

where, $t_{ins}$ is the FE insulator thickness and α, β and γ are the material dependant parameters. For, BaTiO$_3$ the values of these parameters are $-1 \times 10^7$ m/F, $-8.9 \times 10^9$ m$^5$/F/Coul$^2$ and $4.5 \times 10^{11}$ m$^9$/F/Coul$^4$ respectively. Now, if we assume that the higher order terms of eqn. (15) is negligible then we can write.

$$V_{ins} = 2\alpha t_{ins} Q \qquad [16]$$

Following the Salahuddin and Datta (6) formalism if we assume the value of ferroelectric Q is the same as depletion charge $Q_{dep}$ in the source region then

$$V_{ins} = 2\alpha t_{ins} Q_{dep} \qquad [17]$$

$$\text{Therefore, } V_{GS} = V_{GI} + 2\alpha t_{ins} Q_{dep} \qquad [18]$$

Since, here α is negative, $V_{ins}$ will be negative too, which directly results in lower value of $V_{GS}$ than that of $V_{GI}$

## Impact of $V_{DS}$

Equations (10)-(12) and (18) determines the current as a function applied gate voltage for the proposed device with infinite or very high drain source voltage because it was derived on the basis of the following two assumptions

1) Energy band bending in the source region is determined only by the applied $V_{GS}$.

2) All the generated carriers by the tunneling process at the source region contributes to the drain current $I_D$.

To incorporate the effect of $V_{DS}$ on drain current, the following assumptions have been appended with the previous ones.

1) Drain source voltage determines the position of the Fermi level in the drain and this Fermi level is assumed to be constant at source and channel region.

2) The electrons can tunnel into an energy level at or above the electron Fermi level not below the Fermi level.

Fig.2 (b) shows the energy band diagram perpendicular to the gate incorporating the effect of $V_{DS}$ by including the constant Fermi level ($E_F$). Since the energy band profile is same along the gate length ($L_{GS}$), the Fermi level $E_F$ crosses everywhere at the same distance from the oxide semiconductor interface. And tunneling below this cross point is no longer allowed which in effect results in a reduction of tunneling zone. A schematic demonstration of these phenomena is exhibited in Fig.2 (b). So, the total band bending $\varphi_s$ can be calculated from equation (12) when

$$\varphi_s < (V_{DS} + \frac{E_g}{q}) \qquad [19]$$

But, in case of large gate bias and low drain-source voltage when the total band bending calculated from equation (12) crosses this limit, the new total band bending should be calculated from the following equation.

$$\varphi_{s,new} = (V_{DS} + \frac{E_g}{q}) \qquad [20]$$

So, after a certain gate voltage the total band bending is pinned at a constant value which in effect results in saturation current.

### Results and Discussions

In this paper all the modeled results of NCTFET is based on $BaTiO_3$ ferroelectric material and flat band voltage $V_{FB}$ is assumed to be zero in all the cases. Fig. 3 exhibits the $I_{DS} - V_{GS}$ plot of NCTFET along with conventional TFET. Careful investigation of this plot reveals two observable facts, first, tunnel current starts to grow at lower $V_{GS}$ and second, the rate of current increment is higher in case of NCTFET than that of conventional TFET. These phenomena is the direct consequence of surface potential amplification by the FE complementary gate oxide. Table I presents a comparison between subthreshold slope parameter of NCTFET and TFET for two different drain currents. From these comparisons it is evident that NCTFET yields a very low SS than that of traditional TFET. The input characteristics of proposed NCTFET with different $V_{DS}$ are shown in Fig.4 (a). Fig. 4 (b) shows the output characteristics of the proposed device. This curve can be interpreted in the following way, with the increment of $V_{DS}$ tunneling zone enhances so thus the drain current but after a certain drain voltage tunnel path reaches its shortest possible length for a fixed source doping then increment of drain voltage does not increase tunnel zone thus drain current saturates.

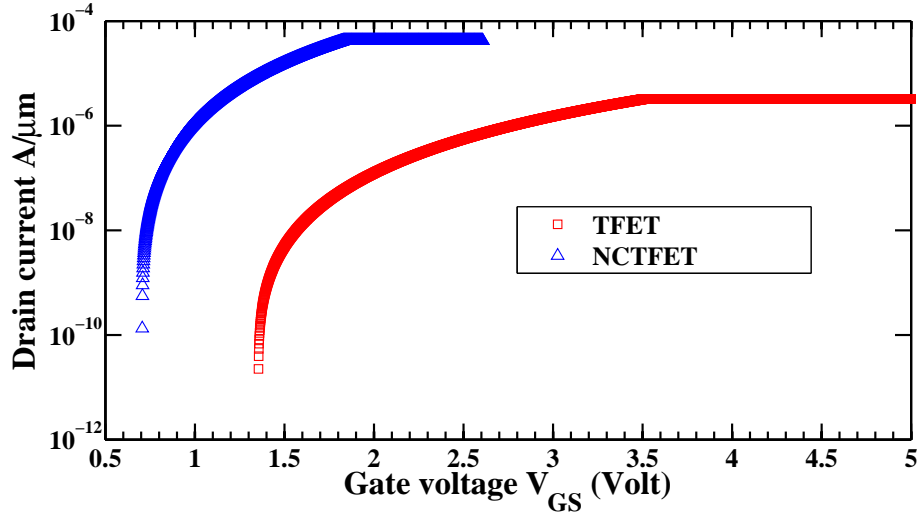

Figure 3. $I_{DS}$–$V_{GS}$ characteristics of NCTFET and conventional TFET (as modeled by Ref.(11))

**TABLE I.** Table for Subthreshold Slope Comparison.

| Drain Current $I_D$ (A/µm) | SS of NCTFET (mV/decade) | SS of TFET (mV/decade) |
|---|---|---|
| $9 \times 10^{-10}$ | 4.83 | 84 |
| $2 \times 10^{-9}$ | 13.78 | 125 |

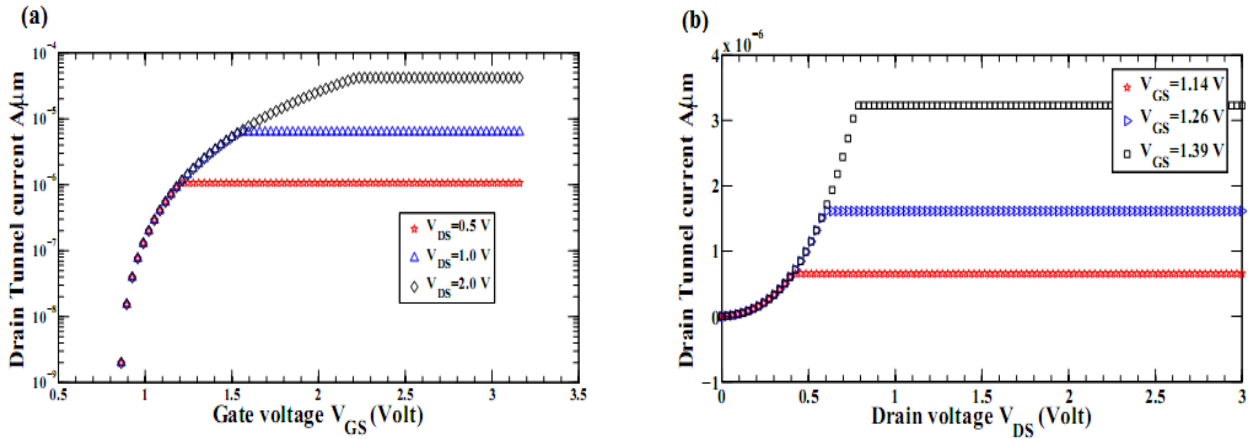

Figure 4: (a) Input characteristics of NCTFET with Source Doping Na=$10^{26}$m$^{-3}$ Complementary Gate dielectric (BaTiO$_3$) thickness 300nm and $V_{FB}$ = 0V. (b) Output characteristics of NCTFET with Source Doping Na = $10^{26}$m$^{-3}$ Complementary Gate dielectric (BaTiO$_3$) thickness 300nm and $V_{FB}$ = 0V.

This superlinear behavior is the direct consequence of parabolic bandstructure in the source depletion region. The exponential dependence of ON current on tunnel path length results in a superlinear current increase. Though FE material does not necessarily reduce the detrimental as well as unexpected superlinear behavior but it can be easily reduced by increasing the source doping or by using low bandgap material.

## Conclusion

In this paper a novel device is proposed and an analytical model of the tunnel current is presented. From the modeled characteristics it has been demonstrated that NCTFET has lower SS than the traditional TFET and can provide a high ON current with low operating voltage. A significant feature of the proposed device is that all remarkably improved characteristics have been obtained without sacrificing commercially well-matured silicon substrate. So, the proposed NCTFET certainly can be a potential candidate for the next generation low power transistor operation and beyond.

## References

1. D. B. Tuckerman and R. F. W. Pease, *IEEE Trans. Elec. Dev.*, 28, 1230 (1981).
2. K. Gopalakrishnan, P. B. Griffin, and J. D. Plummer, *IEDM Tech. Dig.*, pp. 289-292, 2002.
3. H. Kam, D. T. Lee, R. T. Howe, and T. J. king, *IEDM Tech. Dig.*, pp 463-466, 2005.
4. N. Abele , N. Fritschi , K. Boucart , F. Casset , P. Ancey and A. M. Ionescu, *IEDM Tech. Dig.*, pp.1075 -1077, 2005.
5. P.-F. Wang, K. Hilsenbeck, T. Nirschl, M. Oswald, C. Stepper, M. Weiss, D. Schmitt-Landsiedel, and W. Hansch, *Solid State Electron.*, vol. 48, no. 12, pp.2281-2286, May 2004.
6. S. Salahuddin and S. Datta, *Nano Letters*, vol.8, no.2, pp. 405-410, 2008.
7. Anne S. Verhulst, Daniele Leonelli, Rita Rooyackers, and Guido Groeseneken, *J. Appl. Phys.* 110, 024510 (2011)
8. Anne S. Verhulst, William G. Vandenberghe, Karen Maex, and Guido Groeseneken, *Appl. Phys. Lett.* 91, 053102 (2007).
9. Vandenberghe, W., Verhulst, A.S., Groeseneken, G., Soree, B., Magnus, W. *in International Conference on Simulation of Semiconductor Processes and Devices, 2008. SISPAD 2008.*, On page(s): 137 - 140
10. G. A. Salvatore, D. Jimnez and A. M. Ionescu, *Proc. IEDM 2010*, page(s):16.3.1 - 16.3.4
11. E. O. Kane, *Journal of Physics and Chemistry of Solids*, vol. 12, pp. 181-188, 1959.
12. L. D. Landau and I. M. Khalatnikov, Dok. Akad. Nauk, *SSSR*, vol. 96, pp. 469-472, Jun, 1954.
13. V. C. Lo, *Journal of Applied Physics*, Vol. 93, No. 5, pp. 3353-3359, 2003.